\documentclass[10pt]{article} 
\usepackage{latexsym}
\usepackage{amsmath}
\usepackage{amssymb}
\usepackage{pstricks}
\usepackage{indentfirst}

\long\def\ca#1\cb{} 

\newcommand{\ad}{^\dagger }

\newcommand{\AND}{{\small AND}}

\newcommand{\becs}{\begin{cases}}
\newcommand{\bem}{\begin{matrix}}
\newcommand{\blp}{\bigl(}

\newcommand{\brp}{\bigr)}

\newcommand{\dya}[1]{|#1\rangle\langle#1|}

\newcommand{\encs}{\end{cases}}
\newcommand{\enm}{\end{matrix}}

\newcommand{\inp}[1]{\langle#1|#1\rangle }
\newcommand{\inpd}[2]{\langle#1|#2\rangle }
\newcommand{\ket}[1]{|#1\rangle }

\newcommand{\lra}{\leftrightarrow }

\newcommand{\mte}[2]{\langle#1|#2|#1\rangle }

\newcommand{\NOT}{{\small NOT}}
\newcommand{\od}{\odot }

\newcommand{\ot}{\otimes }

\newcommand{\ra}{\rightarrow }
\newcommand{\Ra}{\Rightarrow }

\newcommand{\rgl}{\rangle }

\newcommand{\st}{\sqrt{2}}

\newcommand{\Tr}{{\rm Tr}}

\newcommand{\vbB}{\boldsymbol{\mid}}

\newcommand{\EC}{{\mathcal E}}
\newcommand{\FC}{{\mathcal F}}

\newcommand{\HC}{{\mathcal H}}

\newcommand{\MC}{{\mathcal M}}

\newcommand{\PC}{{\mathcal P}}
\newcommand{\QC}{{\mathcal Q}}

\newcommand{\SC}{{\mathcal S}}

\newcommand{\al}{\alpha }

\newcommand{\gm}{\gamma }
\newcommand{\Gm}{\Gamma }
\newcommand{\dl}{\delta }

\def\outl#1{\par{\medskip\noindent\hspace*{.5cm}\bf
      \mathversion{bold}#1\mathversion{normal}\smallskip} }
 \def\xa{} \def\xb{}  

 \def\outl#1{}  \def\xa{} \def\xb{}  

\ca
 \def\outl#1{\par{
\noindent\hspace*{.5cm}\bf
      \mathversion{bold}#1\mathversion{normal}\smallskip} }
 \long\def\xa#1\xb{}
 
\cb

\begin{document}

\title{Epistemic Restrictions in Hilbert Space Quantum Mechanics}

\author{Robert B. Griffiths 
\thanks{Electronic mail: rgrif@cmu.edu}\\  
Department of Physics,
Carnegie-Mellon University,\\
Pittsburgh, PA 15213, USA}
\date{Version of 16 October 2013}
\maketitle  

\xb

\xa
\begin{abstract}
  A resolution of the quantum measurement problem(s) using the
  consistent histories interpretation yields in a rather natural way a
  restriction on what an observer can know about a quantum system, one that is
  also consistent with some results in quantum information
  theory.  This analysis provides a quantum mechanical understanding of
   some recent work that shows that certain kinds of quantum
  behavior are exhibited by a fully classical model if by hypothesis an
  observer's knowledge of its state is appropriately limited.
\end{abstract}

\xb
\xa
\xb
\section{Introduction}
\label{sct1}
\xa

\xb
\outl{Two strategies to better understand QT, (i) start from CM; (ii) start
  from QM}
\xa

The problem of understanding the quantum world continues to give rise to
numerous debates.  While the tools of textbook quantum theory allow us to
calculate probabilities of measurement outcomes in agreement with experiment,
the problem of understanding these in terms of microscopic quantum phenomena
continues to perplex beginning students as well as their teachers. There are
(at least) two distinct strategies for exploring these questions.  One starts
with classical physics, which is reasonably well understood, both its
mathematical structure and its physical or intuitive interpretation, and tries
to see how far classical ideas can be pushed into the quantum domain before
they fail.  This helps locate the classical-quantum boundary, and identify
which classical concepts remain useful once it has been crossed, and which must
be abandoned or radically modified.  A second strategy starts from a consistent
formulation of microscopic quantum theory, and seeks to apply it to larger
systems to see how classical physics emerges as a suitable, and sometimes
extremely good, approximation to quantum theory at the macroscopic level.

\xb
\outl{First strategy: HV as per Bell. Useful approach despite some dubious
  claims }
\xa

Much current research on hidden variable models represents the first strategy.
In a version pioneered by John Bell, one starts with hypotheses which seem
plausible in classical physics and uses them to deduce consequences, typically
inequalities, whose violation by quantum theory and experiment shows that one
or more of the assumptions made in the derivation do not apply to the real
quantum world.  While some of the resulting claims, such as that the quantum
world is nonlocal or contextual, do not stand up under scrutiny \cite{Grff11,
Grff13b,Grff13c}, this research should nonetheless help us better
understand quantum mysteries provided the classical ideas and assumptions
underlying the hidden variables approach are clearly and properly identified.

\xb
\outl{Spekkens' toy model; BRS Cl harmonic oscillators}
\xa

Classical ideas are made quite explicit in Spekkens ``toy theory'' approach
\cite{Spkk07}, where by hypothesis an observer can have only a limited
knowledge of the actual (ontic) state represented by some collection of
classical variables. This idea has recently been extended in a very careful
study \cite{BrRS12} of coupled classical harmonic oscillators, by assuming that
an observer's knowledge, in the form of a probability distribution, is limited
by an \emph{epistemic restriction} that resembles a quantum uncertainty
principle.  This restriction allows the authors to reproduce in an explicitly
classical model a number of ``weird'' effects previously thought to lie wholly
in the domain of quantum physics.  To be sure, this approach does not reproduce
the entire gamut of quantum phenomena, but the results encourage the authors to
believe, as stated in their introduction, that there might be an axiomatization
of quantum theory in which the first axiom states a fundamental restriction on
how much observers can know about a system, and the second embodies some novel
principle about quantum reality (rather than knowledge thereof).  They then
add, ``Ultimately, the first axiom ought to be derivable from the second
because what one physical system can know about another ought to be a
consequence of the nature of the dynamical laws.''

We shall show that this is not a vain hope. The ``novel principle'' has already
appeared in the physics literature as part of an approach embodying the second
strategy mentioned above, the effort to understand how classical physics is an
approximation to a more exact underlying quantum theory when the latter is
properly understood and interpreted.  What is known as the ``consistent'' or
``decoherent'' histories---hereafter referred to simply as
``histories''---program, introduced in \cite{Grff84,Omns88,GMHr90}, provides,
on the one hand, a fully consistent and paradox free (so far as is known at
present) approach to microscopic quantum phenomena, and on the other a means
for showing that the laws of classical mechanics are in appropriate
circumstances a good approximation to the underlying and more exact quantum
physics.  In particular the \emph{single framework rule} of Hilbert space
quantum mechanics is a novel principle (relative to classical physics) that
leads in a rather natural way to an epistemic restriction of a quite
fundamental sort: what an observer can know is limited by the nature of quantum
reality, since that which does not exist also cannot be known.

The remainder of this paper is organized as follows. Since no literature
pertaining to the histories approach is as much as mentioned in
\cite{Spkk07,BrRS12}, Sec.~\ref{sct2} contains a brief summary of the relevant
principles, including the single framework rule.  For additional details we
refer the reader to other summaries as well as more extensive treatments of the
basic ideas; the following are listed in order of increasing length:
\cite{Grff09b,Hhnb10,Grff11b,Grff13,Grff02c}.  In Sec.~\ref{sct3} we show how
these principles resolve the measurement problem(s) of quantum foundations,
leading in a rather natural way to restrictions on what can be learned using
measurements. Section~\ref{sct4} argues that the results of Sec.~\ref{sct3} are
consistent with quantum information theory. The results are summarized in the
concluding Sec.~\ref{sct5}.

\xb
\section{Hilbert Space Quantum Mechanics}
\label{sct2}
\xa

\xb
\subsection{Quantum properties}
\label{sbct2.1}
\xa

\xb
\outl{Property $\lra$ subspace; spectral representation of observable $A$
using $\{P_j\}$}
\xa

A key idea that goes back to von Neumann, Ch.~III of \cite{vNmn32b}, is that a
\emph{physical property}---something which can be true or false, such as ``the
energy is between $2$ and $3$ J''---is represented in quantum mechanics by a
(closed) \emph{subspace} of the quantum Hilbert space or, equivalently, by the
\emph{projector} (orthogonal projection operator) onto this subspace.  (Here
and later we assume a finite-dimensional Hilbert space; infinite dimensions
complicates the mathematics without resolving any of the quantum conceptual
difficulties.)  A \emph{physical variable} or \emph{observable}, such as energy
or angular momentum, is represented by a Hermitian operator which can be
written in the form
\begin{equation}
A = \sum a_j P_j,\quad P_j =  P_j\ad = P_j^2,\quad \sum_j P_j = I.
\label{eqn1}
\end{equation}
Here the $\{P_j\}$ are a collection of projectors that form a (projective)
\emph{decomposition of the identity} operator $I$, and the $\{a_j\}$, each of
which occurs but once in the sum, so $j\neq k$ means $a_j\neq a_k$, are the
eigenvalues of $A$.  The property that the physical variable $A$ takes on or
possesses the value $a_j$, thus $A=a_j$, corresponds to the projector $P_j$ or,
equivalently, the subspace $\PC_j$ onto which $P_j$ projects.

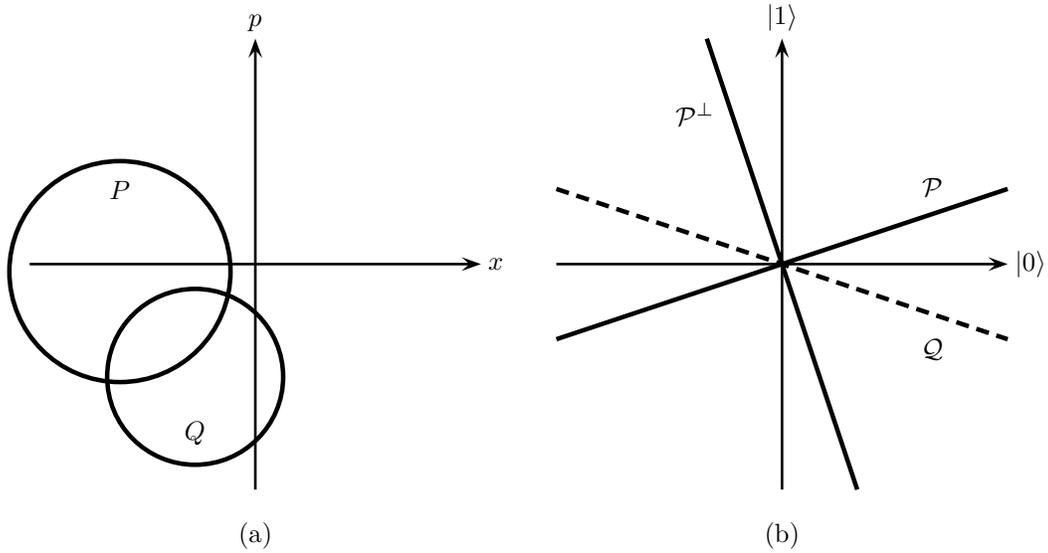
\begin{figure}[h]
$$
\begin{pspicture}(-7,-4)(7,3.5) 
\newpsobject{showgrid}{psgrid}{subgriddiv=1,griddots=10,gridlabels=6pt}
\def\lwd{0.035} 
\def\lwb{0.06}  
\def\rdot{0.13} \def\rodot{0.2} 
\psset{
labelsep=2.0,
arrowsize=0.150 1,linewidth=\lwd}
\def\dot{\pscircle*(0,0){\rdot}} 
\def\odot{\pscircle[fillcolor=white,fillstyle=solid](0,0){\rodot}} 
		\def\pspace{
\psline{->}(-3,0)(3,0)
\psline{->}(0,-3)(0,3)
\rput[l](3.1,0){$x$}
\rput[b](0,3.1){$p$}
\pscircle[linewidth=\lwb](-1.8,-0.1){1.5}
\pscircle[linewidth=\lwb](-0.8,-1.5){1.2}
\rput[t](-1.8,1.1){$P$}
\rput[b](-0.8,-2.4){$Q$}
\rput[B](0,-3.7){(a)}
		}
		\def\hspace{
\psline{->}(-3,0)(3,0)
\psline{->}(0,-3)(0,3)
\rput[l](3.1,0){$|0\rgl$}
\rput[b](0,3.1){$|1\rgl$}
\psline[linewidth=\lwb](-3,-1)(3,1)
\psline[linewidth=\lwb](-1,3)(1,-3)
\psline[linewidth=\lwb,linestyle=dashed](-3,1)(3,-1)
\rput[b](2,0.9){$\PC$}
\rput[r](-0.9,2){$\PC^\perp$}
\rput[t](2,-1){$\QC$}
\rput[B](0,-3.7){(b)}
		}
\rput(3.5,0){\hspace}
\rput(-3.5,0){\pspace}
\end{pspicture}
$$
\caption{(a). Phase space $\Gm$. Properties are indicated by collections of
  points such as $\PC$ and $\QC$. (b) Hilbert space with properties represented
  by rays such as $\PC$, $\PC^\perp$,  and $\QC$. }
\label{fgr1}
\end{figure}
\xb
\outl{Sets in classical phase space; indicators; set operations \& indicator
  algebra}
\xa

In classical mechanics a property corresponds to a set of points $\PC$ in a
classical phase space $\Gm$, as in Fig.~\ref{fgr1}(a), and the classical
counterpart of a projector is an \emph{indicator function} $P(\gm)$ on the
phase space, which takes the value 1 if $\gm$ lies inside $\PC$, and 0 if
$\gm$ lies in the \emph{complement} $\PC^c$ of $\PC$, the points in $\Gm$ that
are not in $\PC$.  Obviously the two descriptions, using the set $\PC$ or the
indicator $P$, are equivalent, and set theoretic operations on sets correspond
to arithmetic operations on indicators. Thus the indicator of the intersection
$\PC\cap\QC$, the property ``$P$ \AND\ $Q$'' or $P\land Q$, is the product
$PQ$, and the indicator for $\PC^c$, the property ``\NOT\ $P$'' or $\lnot P$,
is given by $I-P$, where $I$ is the function on $\Gm$ everywhere equal to 1.

\xb
\outl{Quantum negation using $\PC^\perp \lra I-P$ $\Ra$ revision of rules of
  reasoning, truth}
\xa

In quantum mechanics, again following von Neumann, the negation $\lnot P$ of a
property $P$ is given not by the set theoretic complement $\PC^c$ of the
subspace $\PC$, but instead by its \emph{orthogonal complement} $\PC^\perp$,
the collection of all kets (vectors) which are orthogonal to those in $\PC$.
Indeed, $\PC^c$ is not a subspace, whereas $\PC^\perp$ is a subspace with
projector $I-P$.  The situation is shown schematically for a two-dimensional
Hilbert space in Fig.~\ref{fgr1}(b), where the subspace or ray consisting of
all multiples of some nonzero $\ket{\psi}$ is labeled $\PC$ and its orthogonal
complement is the ray $\PC^\perp$. In the classical case any point that is
outside $\PC$ is inside $\PC^c$, corresponding to the two possibilities that
this property is either true or false. By contrast, in the quantum case there
are rays, such as $\QC$ in Fig.~\ref{fgr1}(b), which are different from both
$\PC$ and $\PC^\perp$.  Thus once one accepts von Neumann's prescription for
quantum properties and their negations, a prescription which lies behind but is
seldom clearly explained in textbook discussions, it is clear that the
move from classical to quantum mechanics must include some changes in ideas
about logical reasoning and truth.

\xb
\outl{Quantum conjunction: intersection of subspaces (B \& vN)$\ra$ paradoxes}
\xa

This becomes even clearer when considering the conjunction, ``$\PC$ \AND\
$\QC$'' or $\PC\land\QC$, of two distinct physical properties.  In the
classical case this corresponds to the intersection $\PC\cap\QC$ of the two
sets, or the product $PQ$ of their indicators.  For the quantum case Birkhoff
and von Neumann \cite{BrvN36} proposed using the intersection of two subspaces,
which is itself a subspace, to represent the conjunction of quantum properties,
with the disjunction, ``$P$ or $Q$ or both'', $P\lor Q$, defined as the span of
the set-theoretic union of $P$ and $Q$, consistent with the usual rule that
$\lnot (P\lor Q) = (\lnot P)\land(\lnot Q)$.  While this seems plausible, the
resulting logical structure promptly leads to paradoxes---see Sec.~4.6 of
\cite{Grff02c} for a simple example---if one employs the usual rules for reasoning
about properties.  Birkhoff and von Neumann were aware of this, and their
remedy was to abandon the distributive law $P\land(Q\lor R)= (P\lor Q)\land
(P\lor R)$ as a rule of reasoning in the scheme of \emph{quantum logic} they
proposed.  Despite a great deal of effort, quantum logic has not turned out to
be a useful tool for understanding quantum mechanics and resolving its
conceptual difficulties \cite{Bccg09,Mdln05}.%
\footnote{This may simply reflect the fact that physicists are not smart
  enough.  I tell my students that perhaps superintelligent robots, when they
  make their appearance, may be able to solve the quantum mysteries using
  quantum logic.  But if they do, will they be able (or even want) to explain
  it to us?} %

\xb
\outl{Histories conjunction $P\land Q$ only defined in $PQ=QP$; otherwise it's
  meaningless}
\xa

The problematic nature of quantum conjunctions is also evident when one uses
projectors. The product $PQ$ of two projectors, corresponding to $P\land Q$, is
itself a projector if and only if $PQ=QP$, in which case it projects onto the
intersection $\PC\cap \QC$ of the corresponding subspaces. But if the two
projectors do not commute, neither $PQ$ nor $QP$ is a projector, and there is
no simple relationship between either of them and the projector onto $\PC\cap
\QC$.  In the histories approach, unlike quantum logic, this is dealt with by
introducing a syntactical rule, an instance of the \emph{single framework
  rule}, that says that the conjunction $P\land Q$ is only defined if the
projectors commute; otherwise, when $PQ\neq QP$, the conjunction of these
properties is undefined or meaningless (in the sense that this interpretation
of quantum mechanics assigns it no meaning).  Note the distinction between a
false statement and a meaningless statement, such as $P\land\lor\, Q$ in
ordinary logic.  The negation of a false but meaningful statement is a true
statement, whereas the negation of a meaningless statement is equally
meaningless.

\xb
\outl{Spin-half example, $S_x$ and $S_z$, to illustrate single framework rule}
\xa

To better understand what is and is not implied by the single framework rule
and its relation to textbook quantum mechanics, consider the Hilbert space of a
spin-half particle, with the orthonormal basis $\{\ket{z^+}, \ket{z^-}\}$
corresponding to $S_z= +1/2$ and $-1/2$ in units of $\hbar$, and
the projectors---we use $[\psi]$ for $\dya{\psi}$ if $\ket{\psi}$ is
normalized---$[z^+]$ and $[z^-]$.  These projectors commute; in fact
$[z^+][z^-] = [z^-][z^+] = 0$, the zero operator, which in quantum mechanics
represents the proposition that is always false.  On the other hand, $S_x=
+1/2$ and $-1/2$ correspond to the projectors $[x^+]$ and $[x^-]$, projecting
on the rays containing $\ket{x^\pm} = (\ket{z^+}\pm\ket{z^-}/\st$, neither of
which commutes with either $[z^+]$ or $[z^-]$.  The single framework rule says
that it is meaningless to simultaneously assign values to $S_x$ and $S_z$,
e.g., ``$S_z = +1/2$ \AND\ $S_x = -1/2$'' makes no sense. Note that the single
framework rule does not at all forbid a quantum discussion or description using
either $S_x$ or $S_z$, both of which could be individually meaningful or
useful; what it says is that the \emph{combination} lacks physical
meaning.

\xb
\outl{Absence of $[z^+]\land [x^-]$ from Hilbert space $\lra$ impossibility
of simultaneous measurement}
\xa

One way to see that the combination ``$S_z = +1/2$ \AND\ $S_x = -1/2$'' cannot
be defined is to note that \emph{every} ray in a (complex) two-dimensional
Hilbert space can be interpreted to mean that the some component of spin
angular momentum, corresponding to some direction in space, has the value
$+1/2$.  There are no rays left over, thus no room in the Hilbert space, for a
property representing such a (supposed) conjunction.%
\footnote{Quantum logic gets around this by assigning to $(S_z = +1/2)\land
  (S_x = -1/2)$ the 0 projector, the property that is always false.  The
  difficulties this leads to are discussed in Sec.~4.6 of \cite{Grff02c}.} %
That it does not make sense is also implicit in the assertion found in
textbooks that there is no way to simultaneously \emph{measure} $S_x$ and
$S_z$.  This is correct, and we shall say more about measurements in
Sec.~\ref{sct3}. However, students would be less confused were they given the
fundamental reason behind this: it is impossible to measure what is not there.

\xb
\outl{Single framework rule resolves many Qm paradoxes}
\xa

In histories quantum mechanics the single framework rule is a basic tool
for resolving all manner of quantum paradoxes, or at least taming them in the
sense of changing them from unresolved conceptual difficulties into
interesting examples of how the quantum world differs from that of everyday
experience. The reader will find numerous examples in Chs.~19 to 25 of
\cite{Grff02c}.

\xb
\subsection{Probabilities}
\label{sbct2.2}
\xa

\xb
\outl{CH use standard probabilities: sample space $\SC$,
event algebra $\EC$, probability measure $\MC$}
\xa

\xb
\outl{Analogy of coarse-grained Cl phase space using cells}
\xa

The standard (Kolmogorov) probability theory used in the histories approach
requires three things: a \emph{sample space} $\SC$ of mutually exclusive
possibilities, an \emph{event algebra} $\EC$, and a probability measure $\MC$
that assigns probabilities to the elements of $\EC$.
Classical statistical mechanics employs the phase space $\Gm$ as the sample
space. However, a more useful analogy for discussing the quantum case is that
of a coarse graining of $\Gm$ formed by dividing it up into a finite number of
nonoverlapping cells which together cover the entire space.  With cell $j$ is
associated an indicator function $P_j(\gm)$ equal to 1 if the point $\gm$ lies
in cell $j$, and 0 otherwise.  Since the cells do not overlap, the product
$P_jP_k$ of two indicators is 0 if $j\neq k$. This corresponds to the fact that
the different cells which form the sample space represent \emph{mutually
  exclusive} possibilities.  In addition, $\sum_j P_j = I$ corresponds to the
fact that at any given time the phase point $\gm$ representing the system must
be in one of the cells; thus one and only one of these mutually exclusive
possibilities is true.  The simplest choice for the event algebra $\EC$ is
the collection of all subsets of $\Gm$ formed by unions of some of the cells
which make up the sample space, with the indicator function of the union equal
to the sum of the indicators of the cells of which it is composed.  Including
the empty set, whose indicator 0 is the function everywhere equal to zero,
results in a Boolean algebra in that the negation $I-P$ of any $P\in \EC$ and
the conjunction $PQ$ of any two of its elements are also members of $\EC$.

\xb
\outl{QM uses PD $\{P_j\}$ for $\SC$, sums for $\EC$ = framework.}
\xa

\xb
\outl{Example: $A=a_j$ as in \eqref{eqn1}}
\xa

By analogy, in the quantum case a sample space $\SC$ is obtained by choosing a
collection of projectors $\{P_j\}$ which sum to the identity $I$---which
implies that the projectors are orthogonal to each other, $P_j P_k = 0 =P_k
P_j$ for $j\neq k$--- as a quantum sample space $\SC$. The set of all
projectors formed by taking sums of some of the projectors in $\{P_j\}$, plus
the 0 operator, is the corresponding quantum event algebra $\EC$. The event
algebra is called a \emph{framework}, a term also used for the projective
decomposition that generates it.  (As there is a one-to-one correspondence
between $\SC$ and $\EC$, this double usage should not cause confusion.)  The
same physical interpretation can be used as in the classical case: the
$\{P_j\}$ constitute a collection of mutually exclusive properties, one and
only one of which is true at a particular time. Thus in the sample space
employed in \eqref{eqn1}, the observable $A$ will possess one and only one of
its eigenvalues.  The event algebra allows more general things;
e.g., ``$A$ has either the value $a_2$ or $a_3$'' is represented by $P_2+ P_3$.

\xb
\outl{Compatible and incompatible Qm frameworks}
\xa

An important difference between the classical and the quantum case is that in
the former if one uses two different coarse grainings, two different
collections of cells, each of which covers the entire phase space, there is
always a common refinement, a coarse graining using cells made up of
intersections of cells from the two collections.  Its event algebra includes
among its members all the members of the event algebras of the two coarse
grainings from which it is derived.  Exactly the same is possible in the
quantum case \emph{if and only if} each of the projectors in one decomposition
commutes with every projector in the other; that is, if the two decompositions
are \emph{compatible}.  Otherwise there is no common refinement, and the single
framework rule prevents putting the frameworks together in a common
probabilistic model.  For example, with specific reference to the observable
$A$ in \eqref{eqn1}, let $Q$ be a projector that does not commute with one of
the $P_j$.  Then the framework $\{Q,I-Q\}$ is incompatible with $\{P_j\}$, and
according to the single framework rule the question ``What is the value of $A$
given that the quantum system has the property $Q$?'' has no meaning.  (The
situation is different if $Q$ is understood as a pre-probability; e.g., the
role played by $[\Psi_2]$ in \eqref{eqn12} in Sec.~\ref{sbct3.3}.)

\xb
\outl{Single framework rule. Liberty, Equality, Incompatibility, Utility}
\xa

Since the single framework rule lacks any exact classical analog, it is easily
misunderstood. The following principles may help prevent such
misunderstanding.  First, the single framework rule allows the physicist
perfect Liberty to construct different, perhaps incompatible, frameworks when
analyzing and describing a quantum system.  No law of nature singles out a
particular quantum framework as the ``correct'' description of a quantum
system; there is, from a fundamental point of view, perfect Equality among
different possibilities.  The key principle of Incompatibility prohibits
\emph{combining} incompatible frameworks into a single description, or
employing them for a single logical argument leading from premisses to
conclusions.  Finally comes Utility: not every framework is useful for
understanding a particular physical situation. It is also important to avoid
the mistake of thinking that the physicist's choice of framework somehow
influences reality.  Instead, quantum reality allows a variety of alternative
descriptions, useful for different purposes, which when they are incompatible
cannot be combined. Different coarse grainings of a classical phase space, or
different views of a mountain from the north and from the south, are classical
analogies which may help in understanding Liberty, Equality, and Utility. But
Incompatibility requires a quantum example, as provided by the $S_x$ and $S_z$
descriptions of a spin-half particle discussed in Sec.~\ref{sbct2.1} above.

\xb
\outl{Probabilities are positive, additive, but are not further
  constrained}
\xa

The probability measure $\MC$ in standard probability theory is a nonnegative
function $\mu$ on the event algebra $\EC$. It is additive over disjoint sets,
and normalized, $\mu(I)=1$.  For our purposes it suffices to assume that a
nonnegative number $\mu_j$ is attached to each indicator $P_j$ in the sample
space in such a way that $\sum_j \mu_j=1$.  From this the probability of
elements of the event algebra is determined in the usual way, e.g.
$\Pr(P_2+P_3)= \mu_2 + \mu_3$. The same procedure works in the quantum case: to
each projector $P_j$ in the decomposition of the identity (quantum sample
space) under consideration one assigns a probability $\mu_j \geq 0$ satisfying
$\sum_j \mu_j=1$ and then sums of these to the projectors making up the
corresponding event algebra.  It is important to note that aside from
positivity, additivity, and normalization, mathematical probability theory
imposes no restrictions on the $\mu_j$.  The same is true for quantum theory,
except that under certain conditions one can use the Born rule and its
extensions in order to generate probabilities for a closed system, as discussed
below.

\xb
\outl{Framework dependence of `true'}
\xa

Probability theory can be understood as an extension of propositional logic,
where probability 1 corresponds to a proposition that is true, and probability
0 to one that is false. In order to maintain the same connection in quantum
theory, it follows that ``true'' and ``false'' must, like probabilities, be
framework-dependent concepts.  This dependence has sometimes been thought to
imply that the histories approach  leads to contradictions, propositions
which are both true and false \cite{Knt97,BsGh99,BsGh00}.  However, the
single framework rule prevents contradictions from arising \cite{GrHr98,
Knt98b,Grff00,BsGh00b,Grff00b,BsGh00c}, and one can show, see Ch.~16
of \cite{Grff02c}, that the histories approach provides a consistent scheme for
probabilistic inference.

\xb
\subsection{Time development}
\label{sbct2.3}
\xa

\xb
\outl{von Neumann and wave function collapse. In CH Qm dynamics is
  \emph{always} stochastic}
\xa

In textbook quantum mechanics Schr\"odinger's equation provides a deterministic
unitary time development of ``the wave function'' until an external measurement
causes a mysterious wave function collapse.  This approach, found in
\cite{vNmn32b}, is widely (and properly) regarded as unsatisfactory.
In the histories approach quantum dynamics is \emph{always} a stochastic
process, whether or not a measurement occurs, and solutions to Schr\"odinger's
equation are used to compute probabilities by means of the Born rule and its
extensions.  Here we summarize the essentials needed for the discussion of
measurements in Sec.~\ref{sct3}.

\xb
\outl{History Hilbert space, history projector $\lra$ sequence of events in
  time}
\xa

Quantum stochastic time development can be described using a \emph{history
  Hilbert space} $\breve\HC$, which for a sequence of events at times $t_0 <
t_1 <\cdots t_f$ is a tensor product
\begin{equation}
\breve\HC = \HC\od\HC\od\cdots \HC,
\label{eqn2}
\end{equation}
of $f+1$ copies of the Hilbert space $\HC$ used for the system at a single
time, where the customary tensor product symbol $\ot$ has been replaced by
$\od$ as a matter of convenience, to have a distinctive symbol
separating events at different times. An individual quantum history of the
simplest sort is a tensor product of projectors
\begin{equation}
 Y = F_0\od F_1\od\cdots F_f,
\label{eqn3}
\end{equation}
and thus itself a projector on the history Hilbert space. Its physical
interpretation is ``property $F_0$ at time $t_0$, then property $F_1$ at time
$t_1$, then \dots'', where ``then'' could be replaced by ``and.''  In
general, successive events are \emph{not} connected with each other in any way
related to Schr\"odinger's equation.

\xb
\outl{Initial pure state; families based on decomposition $\{P_m^{\al_m}\}$ at
  $t_m$}
\xa

Rather than the most general case we restrict ourselves to the situation in
which at time $t_0$ the projector $F_0 = [\Psi_0]$ projects onto a specific
initial state $\ket{\Psi_0}$, and at each later time $t_m$, $F_m$ belongs to
the event algebra generated by a specific decomposition $\{P_m^{\al_m}\}$ of
the identity, $\sum_{\al_m} P_m^{\al_m}=I$. Here the $\al_m$ are labels, not
exponents, the subscript $m$ indicates the time, and different decompositions
may be used at different times. The sample space of histories corresponds to a
collection $\{Y^\al\}$, where $\al = (\al_1,\ldots \al_f)$ is a vector of
labels, and
\begin{equation}
 Y^\al = [\Psi_0]\od P_1^{\al_1}\od P_2^{\al_2}\od \cdots  P_f^{\al_f}.
\label{eqn4}
\end{equation}
If in addition one includes the special history $(I-[\Psi_0])\od I\od I
\od\cdots I$, which is assigned a probability of 0 and hence plays no role in
the following discussion, the history projectors in the sample space sum to
the history space identity $\breve I = I\od I\od \cdots I$, and thus constitute
a set of mutually exclusive possibilities, one and only one of which can be
said to occur.  The collection of all projectors which are sums of some of the
$Y^\al$ forms the event algebra.

\xb
\outl{Time development operator, chain kets, consistency condition,
  probabilities}
\xa

For a closed system that does not interact with an external environment,
solving Schr\"odinger's equation yields a unitary time development operator
$T(t,t')$ for the time interval from $t'$ to $t$; it is equal to
$\exp[-i(t-t')H/\hbar]$ in the case of a time-independent Hamiltonian $H$.
Using this time development operator we define a \emph{chain ket}
\begin{equation}
 \ket{\al} = P_f^{\al_f} T(t_f,t_{f-1}) P_{f-1}^{\al_{f-1}}
 T(t_{f-1},t_{f-2})\cdots
 P_1^{\al_1} T(t_1,t_0) \ket{\Psi_0}
\label{eqn5}
\end{equation}
for every history $Y^\al$ in the sample space.  A family of histories satisfies
the \emph{consistency condition}, and is called a \emph{consistent family},
provided the inner product of two chain kets for distinct elements of the
history sample space vanishes,
\begin{equation}
 \inpd{\al}{\al'} = \mu_\al \dl(\al,\al'),
\label{eqn6}
\end{equation}
where $\dl(\al,\al')$ is 1 if $\al^{}_m=\al'_m$ for every $m$, and is 0
otherwise. When \eqref{eqn6} is satisfied the $\mu_\al$ are the (extended) Born
probabilities for histories in the sample space, and determine the
probabilities for histories in the event algebra in the usual way.
Condition \eqref{eqn6} is needed to ensure that the probabilities defined in
this way satisfy the usual rules of probability theory; e.g., if one sums the
joint distribution over all possibilities at a particular time $t_j$ the result
should be the joint distribution for the events at the remaining times as
calculated omitting all mention of $t_j$ from the family of histories.  For a
more detailed discussion of this point see Sec.~10.2 of \cite{Grff02c}.

\xb
\outl{Born rule as special case. Probabilities refer to closed system, not
  measurements}
\xa

In the case $f=1$, histories involving only two times $t_0$ and $t_1$, the
consistency condition is automatically satisfied, since the projectors at time
$t_1$ are orthogonal to each other, and the probabilities are exactly those
given by the usual Born rule.  Note, however, that these probabilities refer to
states of affairs inside a closed quantum system, not to outcomes of
measurements carried out on that system by some external apparatus.  The
overall consistency of this approach is shown in Sec.~\ref{sct3} below,
where measurements themselves are treated as quantum mechanical processes
occurring within a (large) closed system.

\xb
\outl{Consistency condition is restrictive, $\ra$ extension of single framework
rule}
\xa

When $f=2$ or more, a family of histories involving three or more times, the
consistency condition \eqref{eqn6} on the orthogonality of chain kets for
$\al\neq\al'$ is quite restrictive.  Families of histories for which it is not
satisfied cannot be assigned probabilities in a consistent manner.  It may be
that even when the history projectors of two different consistent families
commute with each other, so that there is a common refinement, this refinement
does not satisfy the consistency conditions, and so cannot be assigned
probabilities.  It is then natural to extend the single framework rule to
include a prohibition of such combinations.%
\footnote{ The three-box paradox discussed in Sec.~22.5 of \cite{Grff02c}
  provides a simple example.} %

\xb
\outl{Various extensions, such as initial density operator, mentioned}
\xa

There are various extensions of the type of analysis given above to more
general situations. In place of an initial pure state $[\Psi_0]$ at $t_0$ one
can use a more general projector or a density operator; in that case chain
operators are employed in place of chain kets.  Sometimes the weaker
requirement that the real part of $\inpd{\al}{\al'}$ vanish for $\al\neq\al'$,
allowing the imaginary part to be nonzero, is used in place of \eqref{eqn6},
though there are reasons \cite{Dsi04} for preferring the stronger condition.%
\footnote{Whereas this weaker requirement is mentioned in Ch.~10 of
  \cite{Grff02c}, all the examples in that book conform to the stronger
  condition, i.e., \eqref{eqn6}.} %
Rather than assuming an initial state at $t_0$ one can use a final state at
$t_f$, or indeed some property at an intermediate time, as the ``initial''
condition.  In constructing a history family the choice of projectors at a
particular time can be made dependent on which event occurred at an earlier (or
at a later) time. Numerous examples illustrating some of these points will be
found in \cite{Grff02c}.  (Using the Heisenberg representation for projectors
that enter chain kets or chain operators results in more compact expressions
that lead to the same consistency conditions and probabilities as in the more
intuitive Schr\"odinger picture employed here.)

\xb
\section{Measurements}
\label{sct3}
\xa

\xb
\subsection{Two measurement problems}
\label{sbct3.1}
\xa

\xb
\outl{Knowledge of Qm world requires resolving two  measurement problems}
\xa

\xb
\outl{Quantum principles developed in Sec.~\ref{sct2} will solve both problems,
and show how QM itself gives rise to a natural epistemic restriction}
\xa

It is clear that any claim to know something about a microscopic quantum system
must go beyond elementary human sense impressions and make use of data provided
by suitable instruments that amplify quantum effects and, so to speak, make
them visible; in particular we need to understand physical measurements as
genuine quantum processes. But this is the infamous \emph{measurement problem}
of quantum foundations, which has two parts. The \emph{first measurement
  problem} is that of understanding the macroscopic outcome---we adopt the
picturesque though outdated language of the position of a visible pointer---in
proper quantum mechanical terms.  The \emph{second measurement problem} is to
relate the pointer position to the prior microscopic property the instrument
was designed to measure.  Here 'prior' means \emph{earlier in time}, since very
often the measurement either destroys or radically alters the system being
measured: think of the detection of a gamma ray, or the scattering process by
which it is inferred that a neutrino came from the sun or from a supernova.  We
need a quantum theory of \emph{retrodiction}, inferring something about the
past from present data.  (Not to be confused with \emph{retrocausation}, the
notion that the future can influence the past.)  Obviously, analyzing
measurements as physical processes cannot employ measurement as some sort of
primitive concept, as in textbooks. Hence the need for a fully consistent
description of microscopic quantum properties, one constructed without
using measurement as a primitive concept or axiom, as summarized in
Sec.~\ref{sct2} above.  We shall now show how these principles can be used to
resolve both measurement problems. The result will then be used in
Sec.~\ref{sct4} to argue that Hilbert-space quantum mechanics itself gives
rise, in a rather natural way, to an epistemic restriction which does not need
to be added as an extra axiom.

\xa
\subsection{Quasiclassical frameworks}
\label{3.2}
\xb

\xb
\outl{QM macroscopic description using quasiclassical framework: Gell-Mann \&
  Hartle}
\xa

\xb
\outl{Classical chaos}
\xa

Describing ordinary macroscopic objects in a consistent, fully
quantum-mechanical fashion is a nontrivial problem, and it would be premature
to claim that every detail has been worked out.  Nevertheless, the work of
Omn\`es \cite{Omns99,Omns99b}, and Gell-Mann and Hartle
\cite{GMHr93,GMHr07,Hrtl11} provides a general procedure which seems adequate
to the task.  We will briefly describe the strategy used by Gell-Mann and
Hartle (also see Ch.~26 of \cite{Grff02c}).  The
first idea is that classical properties can be usefully described using a
\emph{quasiclassical} quantum framework employing coarse-grained projectors
that project onto Hilbert subspaces of enormous, albeit finite, dimension,
suitably chosen so as to be counterparts of classical properties such as those
used in macroscopic hydrodynamics.  Next one argues that the stochastic quantum
dynamics associated with a family of histories constructed using these
coarse-grained projectors gives rise, in suitable circumstances, to individual
histories which occur with high probability and are quantum counterparts of the
trajectories in phase space predicted by classical Hamiltonian mechanics. There
are exceptions; for example, in a system whose classical dynamics is chaotic
with sensitive dependence upon initial conditions one does not expect the
quantum histories to be close to deterministic, but then in practice one also
has to replace the deterministic classical description with something
probabilistic in order to obtain useful results.

\xb \outl{Qcl framework not unique; only one needed for Cl physics, so no SFR
  in CP} 
\xa

A quasiclassical family can hardly be unique given the enormous size of the
corresponding Hilbert subspaces, but this is of no great concern provided
classical mechanics is reproduced to a good approximation, in the sense just
discussed, by any of them.  Therefore all discussions which involve nothing but
classical physics can, from the quantum perspective, be carried out using only
a single quasiclassical framework. And as long as reasoning and descriptions
are restricted to this one framework there is no need for the single framework
rule, which explains why a central principle needed to understand quantum
mechanics is completely absent from classical physics.

\xb
\outl{Dowker and Kent objection does not rule out use of Qcl framework}
\xa

Sometimes the objection has been raised \cite{DwKn96,Knt98} that quasiclassical
frameworks are not the unique possibilities allowed by the histories approach
to quantum mechanics.  In particular, a consistent family can be constructed in
which the projectors are quasiclassical up to some time, and then followed by a
completely different type of projector at later times, and there is no reason
from the perspective of fundamental quantum mechanics to disallow
this. However, there is also no reason to prefer it. The histories approach
does not deny that other incompatible consistent families can be constructed;
it simply insists that this possibility does not invalidate a description
employing a quasiclassical framework, which is what is needed for thinking
about pointer positions. See the discussion of Liberty, etc. in
Sec.~\ref{sbct2.2}.  By analogy, the possibility of using $S_x$ to describe a
spin-half particle does not invalidate a description based on $S_z$; what
cannot be done is to combine them.

\xb
\subsection{A measurement model}
\label{sbct3.3}
\xa

\xb
\outl{Measurement model: $\HC_S$=particle, $\HC_M$= apparatus;
initial states $\ket{\psi_0}$, $\ket{\Psi_0}$}
\xa

To see how the histories approach resolves both measurement problems we
consider a simple model of the measuring process that, apart from minor
changes, goes back to Ch.~VI of  \cite{vNmn32b}.
Let $\HC_S$ be the Hilbert space of the system to be measured, henceforth
referred to as a \emph{particle}, and $\HC_M$ that of the measuring device. For
example, $\HC_S$ could be the 2-dimensional Hilbert space of the spin of a
spin-half particle, while the quantum description of its position might be
among the variables included in $\HC_M$. Let $\{\ket{s^j}\}$ be an orthonormal
basis for $\HC_S$, with states labeled by a superscript so that the subscript
position can refer to time. At $t_0$ let $\ket{M_0}$ be the initial
(normalized) state of the apparatus, while
\begin{equation}
\ket{\psi_0} = \sum_j c_j \ket{s^j},\quad
\ket{\Psi_0}=\ket{\psi_0}\ot\ket{M_0} 
\label{eqn7}
\end{equation}
are the initial states of the particle and of the total closed system that
includes both particle and measuring device. The $c_j$ are complex numbers
satisfying $\sum_j|c_j|^2=1$, so both $\ket{\psi_0}$ and $\ket{\Psi_0}$ are
normalized.

\xb
\outl{Unitary time development of model}
\xa

Let $T(t,t')$, as in Sec.~\ref{sbct2.3}, be the unitary time development
operator for the total system, and assume it is trivial, equal to the identity
operator $I=I_S\ot I_M$, for $t$ and $t'$ both less than some $t_1$ or both
greater than $t_2$, and that for the interval from $t_1$ to $t_2$,
\begin{equation}
 T(t_2,t_1) \blp \ket{s^j}\ot\ket{M_0}\brp = \ket{\hat s^j}\ot \ket{M^j}.
\label{eqn8}
\end{equation}
Here the $\{\ket{M^j}\}$ are orthonormal states of the apparatus,
$\inpd{M^j}{M^k}=\dl_{jk}$, corresponding to different pointer positions, and
the $\{\ket{\hat s^j}\}$ on the right side of the equation---note they are
\emph{not} assumed to be the same as the $\{\ket{s^j}\}$ on the left side---are
normalized, $\inp{\hat s^j}=1$, though (unlike von Neumann's original model)
not necessarily orthogonal.  The transformation \eqref{eqn8} is unitary, or, to
be more precise, it can be extended to a unitary transformation on
$\HC_S\ot\HC_M$, because the orthogonality of the $\ket{M^j}$ ensures that the
states on the right side of \eqref{eqn8} are mutually orthogonal, even though
this may not be true of the $\{\ket{\hat s^j}\}$.  Noting that $T(t_2,t_0) =
T(t_2,t_1)\cdot I$ and applying \eqref{eqn8} to \eqref{eqn7}, one sees that
unitary time development leads to states
\begin{align}
\ket{\Psi_1} = T(t_1,t_0) \ket{\Psi_0} &= \ket{\Psi_0},\notag\\
\ket{\Psi_2} = T(t_2,t_0) \ket{\Psi_0} &= \sum_j c_j\ket{\hat s^j}\ot\ket{M^j}
\label{eqn9}
\end{align}
for the total system at times $t_1$ and $t_2$.

\xb
\outl{Family $\FC_0$, unitary time development, cannot describe measurement
  outcomes }
\xa

We now wish to consider various consistent families that begin with
$[\Psi_0]=\dya{\Psi_0}$ at time $t_0$. One
possibility is unitary time development:
\begin{equation}
 \FC_0:\;\;[\Psi_0]\;\od\; \{[\Psi_1],I-[\Psi_1]\}\;\od\; 
\{[\Psi_2], I-[\Psi_2]\}
\label{eqn10}
\end{equation}
for times $t_0<t_1<t_2$, where the different histories in the sample space are
made up by choosing one of the projectors inside the curly brackets at each of
the later times.  Since the events $I-[\Psi_1]$ and $I-[\Psi_2]$ occur with
zero probability they could actually be omitted without causing confusion;
only the history $[\Psi_0]\od [\Psi_1]\od [\Psi_2]$ occurs, and it is assigned
a probability of 1. While $\FC_0$ is perfectly acceptable as a family
of quantum histories, it cannot be used to discuss possible outcomes of the
measurement because it does not include the projectors $\{[M^k]\}$ for the
pointer positions at time $t_2$, and it cannot be refined to include them
because $[\Psi_2]$ does not commute with some of the $[M^k]$, provided
at least two of the $c_j$ in \eqref{eqn7} are nonzero. Thus the first
measurement problem cannot be solved if one insists that all time development
is unitary and not stochastic.

\xb
\outl{Family $\FC_1$ with measurement outcomes}
\xa

The histories approach can solve the first measurement problem by using the
family
\begin{equation}
 \FC_1:\;\;[\Psi_0]\;\od\;[\Psi_1] \;\od\; \{[M^k]\}.
\label{eqn11}
\end{equation}
in place of $\FC_0$. Here the alternative $I-[\Psi_1]$ at $t_1$, which occurs
with zero probability, has been omitted, and we employ the usual physicist's
convention that $[M^k]=\dya{M^k}$ means $I_S\ot[M^k]$ on the full
Hilbert space $\HC_S\ot\HC_M$. An additional projector $R'=I -\sum_k [M^k]$
should be included at the final time in \eqref{eqn11} so that the total sum is
the $I$, but again it is omitted since its probability is zero.  Note that
there is no reference to the later particle states $\ket{\hat s^k}$ in
\eqref{eqn8}; they are in fact irrelevant for discussing the macroscopic
outcomes of the measurement. While $[\Psi_2]$ cannot be one of the properties
at time $t_2$ in family $\FC_1$, see the discussion of $\FC_0$ above, it can
very well be used as a mathematical device (a pre-probability in the
terminology of Sec.~9.4 of \cite{Grff02c}) to calculate probabilities of the
different pointer positions:
\begin{equation}
\Pr([M^k]_2) = \Tr(\;[\Psi_2]\;[M^k]\;) = \mte{\Psi_2}{\,[M^k]\,}.
\label{eqn12}
\end{equation}
Note that this is a perfectly legitimate and consistent ``epistemic'' use of
$\ket{\Psi_2}$, since it, like a probability distribution, provides some
information about the system, even when it does not represent a physical
property.

\xb
\outl{Family $\FC_2$: particle properties at intermediate time}
\xa

\xb
\outl{From pointer position at $t_2$ one can retrodict particle property at
  $t_1$}
\xa

\xb
\outl{Measured properties and later pointer positions have same probability}
\xa

\xb
\outl{Identity of $\ket{\hat s^j}$ and $\ket{s^j}$ needed for a
  \emph{preparation},  not a  \emph{measurement} }
\xa

In order to relate the measurement outcome to a prior property of the measured 
particle and thus solve the second measurement problem, one needs still another
family :
\begin{equation}
 \FC_2:\;\;[\Psi_0]\;\od\; \{ [s^j]\}\;\od\; \{[M^k]\}.
\label{eqn13}
\end{equation}
Here the decomposition of the identity $\{ [s^j]\}$ at $t_1$ refers to
properties of the particle ($[s^j]$ means $[s^j]\ot I_M$ on the full Hilbert
space) without reference to the apparatus.  It is straightforward to show that
the family $\FC_2$ is consistent, and leads to a joint probability
distribution, where the subscripts on $[s^j]$ and $[M^k]$ refer to the time,
\begin{equation}
\Pr(\,[s^j]_1,[M^k]_2) = |c_j|^2 \dl_{jk},
\label{eqn14}
\end{equation}
with marginals given by
\begin{equation}
 \Pr([s^j]_1) = \Pr([M^j]_2) = |c_j|^2.
\label{eqn16}
\end{equation}
If $|c_k|^2 > 0$ the conditional probability for the earlier property
$[s^j]_1$ given the later pointer position $[M^k]_2$ is
\begin{equation}
 \Pr([s^j]_1 \vbB [M^k]_2) = \dl_{jk}.
\label{eqn15}
\end{equation}
That is to say, from the (macroscopic) measurement outcome or pointer position
$k$ at time $t_2$, standard statistical inference allows one to infer, to
retrodict, that the particle had the property $[s^k]$ at the earlier time
$t_1$.  Also note that, \eqref{eqn16}, the probabilities for particle
properties just before the measurement are identical to those of the later
pointer positions.  This is what one would expect for ideal measurements.  It
shows that textbooks (which follow the lead of \cite{vNmn32b}) in which
students are taught to calculate $|c_j|^2$ for the particle alone and then
ascribe the resulting probability to the outcome of a measurement, are not
wrong, just confusing.
It is worth remarking once again that the later particle states, the
$\ket{\hat s^j}$ in \eqref{eqn8}, play no role in the discussion.  In von
Neumann's model these were set equal to the $\ket{s^j}$, and while this is of
course a perfectly legitimate choice, it has given rise to considerable
confusion in that ``measurements'' are often interpreted as involving a
correlation between the pointer position and the \emph{later} particle state,
something that should properly be called a \emph{preparation}, not a
measurement.

\xb
\outl{More realistic measurement models in CQT}
\xa

The measurement model discussed above is in some ways rather artificial.
However, once one understands its basic features it is possible to construct
more realistic models in which the particle states to be measured form a
general (projective) decomposition of $I_S$ not limited to pure states, the
pointer positions are represented not by pure states but by quasiclassical
projectors onto appropriate Hilbert spaces, the initial state of the apparatus
is a quasiclassical projector or a density operator, and allowance is made for
thermodynamic irreversibility in the measurement process.  See the relevant
sections of Ch.~17 in \cite{Grff02c} for details.  None of these extensions alters
the basic conclusions reached on the basis of the simple model used above.

\xb
\section{Information}
\label{sct4}
\xa

\xb
\outl{Epistemic restriction:  at most $\log d$ (Shannon) in $d$-dimensional
  qudit}
\xa

Given a system with a $d$-dimensional Hilbert space, a projective decomposition
of the identity can contain at most $d$ projectors, one for every element of
some orthonormal basis.  And since a probabilistic quantum description must use
only a single framework, the maximum amount of Shannon information that can be
associated with a sample space of a $d$-dimensional quantum system, the Shannon
entropy $H$ if a probability $1/d$ is assigned to each possibility, is $\log
d$.  Thus a qudit of dimension $d$ (using the terminology of quantum
information theory) cannot receive or contain or carry more than $\log_2 d$
bits of information.  This is a fundamental epistemic restriction arising
directly from the Hilbert space structure of quantum mechanics.

\xb
\outl{Alice sends randomly chosen qudit to Bob, who measures it; Holevo bound}
\xa

This restriction is confirmed by, or is at least consistent with, various
results in quantum information theory. 
A common scenario is one in which Alice prepares a qudit of dimension $d$ in a
known quantum state, chosen from a specified set of possibilities according to
some probability distribution, and sends it to Bob, who is allowed to carry out
a generalized measurement (POVM), or perhaps a collective measurement on
several qudits at the same time, with the aim of determining which states Alice
prepared.  We assume the protocol is repeated $N$ times, and each time Alice
records what she prepares. Similarly, Bob records his measurement outcomes.  As
both of their records belong to the quasiclassical world they can be analyzed
using classical (Shannon) information theory, and a well-known bound due to
Holevo (see, e.g., Sec.~12.1.1 of \cite{NlCh00}) shows that the Shannon mutual
information $H(A:B)$ cannot exceed $N\log_2 d$.  This upper bound is actually
achieved if Alice randomly (with equal probability) prepares states
corresponding to some orthonormal basis, and Bob measures in the same basis.
Thus the limitation implied by the analysis of properties and measurements in
Sec.~\ref{sbct3.3} can, with the help of some not altogether trivial
mathematics, be shown to be quite general.  Neither sophisticated encoding
schemes nor the most general of generalized measurements can do any better than
what is implied by the analysis in Secs.~\ref{sct2} and \ref{sct3}: at most
$d^N$ distinct messages, corresponding to $\log_2 d^N = N\log_2 d$ bits, can be
constructed using $N$ letters chosen from an alphabet of size $d$.

\xb
\outl{Could dense coding, teleportation, or \emph{quantum} info violate
  epistemic limit?}
\xa

\xb
\outl{Quantum info in case of perfect qubit channel}
\xa

The reader could be concerned that this epistemic limit might be violated by
schemes for \emph{dense coding}, or is somehow inconsistent with
\emph{teleportation}, or maybe does not apply to \emph{quantum}, in contrast to
classical, information.  Let us briefly discuss each of these, starting with
the last.  Both ``quantum'' and ``classical'' information remain ill-defined
terms in the quantum information literature, despite (or perhaps because of?)
an enormous number of publications. However, one can understand the basic issue
by means of a simple example, a $d=2$ perfect quantum channel, constructed from
a magnetic-field-free pipe into which Alice sends a spin-half particle, which
is measured by Bob when it emerges at the other end.  If Alice prepares the
state $S_w=+1/2$, where $w$ is some specific direction in space: $z$ or $-z$ or
$x$ or whatever, and Bob measures in the $S_w$ basis, the result will always
(probability 1) be $S_w=+1/2$ and not $S_w=-1/2$.  The basis must be specified,
as there is no way to prepare (or measure) a particle with, say, $S_z=+1/2$
\AND\ $S_x = 1/2$.  Consequently this quantum channel with capacity 1 (qu)bit
cannot actually transmit information at a rate greater than a perfect classical
channel that can only transmit quasiclassical states corresponding to a bit
which is either 0 or 1, and always yields the same output as the input.  More
generally, the quantum capacity cannot exceed the classical capacity (see
Secs.~12.3 and 12.4 of \cite{NlCh00} for technical discussions), and talking
about quantum information (whatever it might be!) does not alter the $\log_2 d$
upper limit for one qudit.

\xb
\outl{Error of particle prepared with $S_w=1/2$ carrying precise info about
  $w$. }
\xa

\xb
\outl{Misleading visualization of spin-half as a spinning top}
\xa

One can be misled by the very useful but somewhat dangerous visualization of a
spin-half particle as a little top spinning about a particular axis.  The
direction $w$ which Alice uses to produce the state $S_w=+1/2$ might be
specified very precisely by some macroscopic setting on her apparatus, so it is
tempting to suppose that information about this precise setting, which might
amount to many bits (depending on the precision), is then carried away by the
particle.  But, as noted in Sec.~\ref{sbct2.1}, there is no room in the quantum
Hilbert space for this kind of information, the distinction one might want to
make between $w$ and a direction $w'$ which is nearby, or even between $w$ and
a $w'$ which is perpendicular to it (e.g, $x$ instead of $z$).%
\footnote{The claim sometimes made that the precise information about $w$ is
  ``really'' present in the particle but cannot be measured reminds one of the
  student who, having just failed the examination, tells the professor he
  understood the subject perfectly, but simply could not recall it during the
  test.} %
A less misleading, albeit still classical, visualization is to think of the
$S_w=+1/2$ state as a spinning top whose axis is oriented at random, but with
the constraint that the $w$ component of its angular momentum be positive.

\ca
Teleportation is, effectively, a quantum channel implemented in an indirect
fashion through the use of an entangled state, a classical channel, and
appropriate measurements and other operations.  No part of the process
indicates any violation of the epistemic limit indicated above.  
\cb

\xb
\outl{Dense coding: info present in 2 particles; , does not exceed
epistemic bound}
\xa

\xb
\outl{Teleportation: refer to literature}
\xa

Dense coding is a process by which $d^2$ messages can be transmitted from Alice
to Bob by sending a single $d$-dimensional qudit through a quantum channel,
provided a fully entangled state of two qudits, one in Alice's laboratory and
the other in Bob's, is initially available.  This might seem to violate the
epistemic limit, since $\log_2 d^2 = 2\log_2 d$ is clearly larger than $\log_2
d$.  The solution to this apparent paradox is a proper microscopic analysis of
where information is located at the intermediate time between Alice's
preparation and Bob's measurement \cite{Grff02}, a type of analysis which is
not easy to do using the tools provided in typical textbooks (including those
devoted to quantum information), because they provide no systematic way of
thinking about microscopic quantum properties during the time interval between
the (quasi)classical preparation and the (quasi)classical measurement.  To
discuss the quantum state of the two qudits, which are initially in a
fully-entangled state, one has to use the tensor product of their Hilbert
spaces, of dimension $d^2$. There is then a projective decomposition of its
identity corresponding to an orthonormal basis containing $d^2$ fully-entangled
states, which are used to encode the $d^2$ messages.  Thus there is no
violation of the epistemic limit, for two particles are involved.
We refer the reader to \cite{Grff02} for a similar discussion of
teleportation; once again, all information can be properly accounted for, and
the fact that two uses of a $d$-dimensional classical channel are required to
complete the protocol, does not imply that a qudit or the corresponding quantum
channel has an information capacity of $2\log_2 d$. (Also see the discussion in
\cite{Grff07} for further insight into the need for a double
usage of a classical channel.)

\xb
\outl{Epistemic restrictions in Qm info theory}
\xa

In addition, quantum information theory contains various epistemic restrictions
on the information that can be obtained from a quantum system or transmitted
from one place to another in the form of inequalities, sometimes referred to as
epistemic uncertainty relations.  Because some of them are obtained using
sophisticated mathematics, and tend to be expressed in terms of
(quasiclassical) preparations and measurements, it is not always made clear
that these, too, arise from the fundamental Hilbert space structure of quantum
theory.  So far as is known at present, information inequalities derived in
this way, such as in \cite{CYGG11}, are entirely consistent with
experiment, in contrast to Bell inequalities and the like, which do not agree
with experiment, and whose derivation is based in a fundamental way upon
classical physics (see, e.g., \cite{Grff11}).

\xb
\section{Conclusion}
\label{sct5}
\xa

\xb
\outl{Qm ontology of Hilbert subspaces restricts what observer can know}
\xa

\xb
\outl{What does not exist cannot be known}
\xa

If one assumes following von Neumann, and consistent with textbook quantum
theory where it is not always clearly discussed, that properties of a quantum
system correspond to subspaces of its Hilbert space, then there is a very
natural epistemic restriction on what an external observer can know about a
microscopic quantum system.  The Hilbert space does not contain, has no room
for, combinations of incompatible properties, such as $S_x=+1/2$ \AND\
$S_z=-1/2$ for a spin half particle, and as a consequence these must be
excluded from a consistent quantum ontology, as discussed in detail in
\cite{Grff13}. And of course what does not exist cannot be known; such an
ontological restriction leads automatically to an epistemic restriction.

\xb
\outl{Epistemic restriction already present in textbooks, Qm info theory}
\xa

Quantum textbooks already contain a version of this epistemic restriction.
Students are told that incompatible quantum properties cannot be simultaneously
measured.  However, because textbooks treat measurements as a sort of axiom
which is incapable of further analysis, a black box which cannot be pried open
to see what is going on inside, the nature of this restriction remains clouded
in a dense conceptual fog.  One needs a consistent quantum analysis of
measurements, one capable of resolving both the first and the second quantum
measurement problems, to relate this restriction on measurements to
mathematical properties of the Hilbert space used in quantum mechanics to
represent physical properties.  And, as noted in Sec.~\ref{sct4}, some of the
rigorous inequalities developed by quantum information theorists using the
quantum Hilbert space are also epistemic restrictions.

\xb
\outl{CH provides ways to understand other Qm phenomena mentioned in
  \cite{BrRS12}}
\xa

\xb
\outl{Additional Cl restrictions or additions $lra$ tweaking Bohr quantization}
\xa

In addition to resolving the measurement problems, the histories approach
provides a foundation for understanding all the other strange, i.e.,
nonclassical, quantum phenomena, including those which cannot at present be
obtained from a classical model by adding an epistemic restriction (see the
discussion in Sec.~V of \cite{BrRS12}).  This is because it has a consistent
formulation of the fundamental principles of Hilbert space quantum mechanics,
the principles that underlie the generally accepted calculational techniques
taught in textbooks.  It may be that applying still more restrictions, or
perhaps additions, to classical models will eventually yield the correct
quantum outcomes.  But one can ask whether such a circuitous route, somewhat
analogous to tweaking Bohr's semiclassical quantization condition, would be
worthwhile, given the availability of a more direct path to understanding the
phenomena in question.

\xb
\outl{Cl models useful in seeking limits of CM, but need to be combined with
CH}
\xa

This is not to say that the study of classical models of the sort considered in
\cite{Spkk07,BrRS12} is without value.  It is surely of interest to understand
the limits of classical physics when it is pushed as far as possible into the
quantum domain.  Not least because in the domain where classical physics
functions very well, the quasiclassical regime of macroscopic phenomena, it
provides a much simpler and easier calculational scheme than any full-scale
quantum counterpart.  Who would ever want to compute an earth satellite orbit
starting with a wave function?  However, such studies of classical models will,
we believe, be most effective when combined with a consistent and complete
microscopic quantum theory, one that takes full account of the noncommutation
of quantum projectors, is not dependent upon a vague concept of
``measurement,'' and does not require any additional epistemic restrictions
beyond those implied by the formalism itself.  It is hoped that the work
presented here will contribute to that end.

\section*{Acknowledgments}

This article had its genesis in a very helpful conversation with Robert
Spekkens, who pointed out to the author the significance of \cite{BrRS12}. The
research described here has been supported by the National Science Foundation
through Grant PHY-1068331.
\xb
\xa

\xb\end{document}